\documentstyle[12pt,bibtmp,epsf]{article}
\def\@normalsize{\@setsize\normalsize{12pt}\xiipt\@xiipt
\abovedisplayskip 6\p@ plus3\p@ minus3\p@
\belowdisplayskip \abovedisplayskip
\abovedisplayshortskip  \z@ plus3\p@
\belowdisplayshortskip  4.5\p@ plus2.5\p@ minus2\p@
\let\@listi\@listI}

\begin{document}
\topmargin 0pt
\oddsidemargin 5mm
\begin{flushright}
{\footnotesize Presented at the MRS Fall meeting in Boston, December 1-5,
1997}
\end{flushright}

\begin{flushleft}
{\bf SPIN TUNNELING IN CONDUCTING OXIDES (invited)
} \\
\end{flushleft}
\noindent
\begin{flushleft}
{Alexander BRATKOVSKY }\\ {
Hewlett-Packard Laboratories, 3500 Deer Creek Road,
Palo Alto, CA 94304-1392, alexb@hpl.hp.com
}
\end{flushleft}

\noindent

\begin{abstract}
{
\footnotesize

Direct tunneling in ferromagnetic junctions
is compared with impurity-assisted, surface state assisted, 
and inelastic contributions to a tunneling magnetoresistance (TMR).
Theoretically calculated direct tunneling in
iron group systems leads to about a 30\% change in resistance, 
which is close to experimentally observed values. It is shown 
that the larger observed
values of the TMR might be a result of tunneling involving 
surface polarized states.
We find that tunneling via resonant defect states in the
barrier radically decreases the TMR
(down to 4\% with Fe-based electrodes), 
and a resonant tunnel diode structure
would give a TMR of about 8\%.
With regards to inelastic tunneling,
 magnons and phonons exhibit opposite effects: 
one-magnon emission generally results in spin mixing and,
consequently,
reduces the TMR, whereas phonons are shown to  enhance the TMR. 
The inclusion of both magnons and phonons reasonably explains an unusual 
bias dependence of the TMR.

The model presented here is applied qualitatively 
to half-metallics with 100\%  spin polarization, where 
one-magnon processes are suppressed and
the change in resistance in the absence of
spin-mixing on impurities may be arbitrarily large. Even in the 
case of imperfect
magnetic configurations, the resistance change can be a few 1000 percent.
Examples of half-metallic systems are  CrO$_2$/TiO$_2$ and 
CrO$_2$/RuO$_2$, and an account of their peculiar band structures
is presented. The  implications and relation of these systems
to CMR materials which are nearly half-metallic, are discussed.

}
\end{abstract}

\normalsize

\begin{flushleft}
INTRODUCTION
\end{flushleft}

Tunnel magnetoresistance (TMR) in ferromagnetic junctions, 
first observed more than a decade ago,\cite{jul,maekawa} is of
fundamental interest and potentially applicable to
 magnetic sensors and  memory devices.\cite{tedrow}
This became particularly relevant after it was found
that the TMR for 3$d$ magnetic electrodes reached large
values at room temperature \cite{moodera,miyazaki},
and junctions demonstrated a non-volatile memory effect.
These observations has ignited a world-wide effort towards using this effect
in various applications, with memories and sensors being the 
most natural choices.

A simple model for spin tunneling has been formulated by Julliere
\cite{jul} and further
developed in Refs.\cite{stearns,slon}. 
This model is expected to 
work rather well for iron, cobalt, and nickel based metals,
according to theoretical analysis\cite{stearns} and experiments.\cite{moodera}
However, it disregards important points such as
impurity-assisted and inelastic scattering, tunneling into surface
states, and the reduced effective mass of carriers inside the barrier.
 These effects are important for proper understanding of the behavior of
actual devices, like peculiarities in their $I-V$ curves,
as considered in Ref.\cite{amb_tunn1} and the present paper. 
I shall also discuss a 
couple of {\em half-metallic} systems which should in principle achieve
the ultimate magnetoresistance at room temperatures and low fields.

\newpage
\begin{flushleft}
ELASTIC AND INELASTIC TUNNELING,  MODEL
\end{flushleft}
The model that we will consider below includes a Hamiltonian for
non-interacting conducting spin-split electrons ${\cal H}_0$,
electron-phonon interaction ${\cal H}_{ep}$, and exchange interaction with 
localized $d_l$ electrons ${\cal H}_{x}$, the later giving rise
to the electron-magnon interaction. Impurities will be described 
by a short-range confining potential $V_i$,
\begin{eqnarray}
{\cal H} &=& {\cal H}_0 + {\cal H}_{ep} + {\cal H}_x 
+ {\cal H}_i,\\
{\cal H}_i &=& \sum_{\bf n_i} V_i({\bf r - n_i})\nonumber
\end{eqnarray}
where ${\bf r}$ stands for the coordinate of the electron and
${\bf n_i}$ denotes the impurity sites.

The non-interacting part of the Hamiltonian ${\cal H}$
describes electrons in the ferromagnetic electrodes and insulating barrier
according to the Schr\"odinger equation~\cite{slon}
\begin{equation}
({\cal H}_{00} - {\bf h} \cdot \hat{\bf \sigma})\psi=E\psi,
\label{eq:uSh}
\end{equation}
where ${\cal H}_{00} = -(\hbar^2/2m_\alpha)\nabla^2+U_\alpha$ is the
single-particle Hamiltonian with $U({\bf r})$ the potential energy,
${\bf h}(\bf r)$ the exchange energy ($=0$ inside the barrier), 
${\bf \sigma}$ stands for the Pauli matrices; 
indices $\alpha$=1, 2, and 3 mark the quantities for 
left terminal, barrier, and right terminal, respectively
(${\cal H}_0$ is the expression in brackets).
We shall also use the following notations to clearly distinguish between
left and right terminal: ${\bf p = k_1}$ and ${\bf k=k_3}$.
Solution to this problem in the limit of a thick barrier provides us 
with the basis functions for
electrons in the terminals and barrier to be used in Bardeen's
tunneling
Hamiltonian approach.\cite{bardeen,mahan} We assume that all many-body
interactions in the electrodes
are included in the effective parameters of (\ref{eq:uSh}).
To fully characterize tunneling we add to Bardeen's term ${\cal
H}_T^0$
the contributions from ${\cal H}_x$ and ${\cal H}_{ep}$:
\begin{eqnarray}
{\cal H}_T &=& {\cal H}^0_T + {\cal H}_T^{x} + {\cal H}_T^{ep},\\
{\cal H}^0_T &=& \sum_{{\bf p,k}a} T^0_{{\bf p}a,{\bf k}a}r^\dagger_{{\bf k}a}
l_{{\bf p}a} + h.c.,\nonumber\\
 T^0_{{\bf p}a,{\bf k}a}&=&{-\hbar^2/(2m_2)}\int_{\Sigma} d{\bf A}\left(
\bar{\psi}_{{\bf k}a}~\nabla{\psi}_{{\bf p}a} - 
\nabla\bar{\psi}_{{\bf k}a}~\psi_{{\bf p}a}\right);
\label{eq:bardeen}\\ 
{\cal H}_T^x
&=& 
- \sum_{\alpha\bf n,k,p} T^{J,\alpha}_{\bf k,p}({\bf n})
\left[(S^3_{\bf n} - \langle S^3_{\bf n} \rangle)  
(r^\dagger_{\bf k\uparrow}l_{\bf p\uparrow} -
r^\dagger_{\bf k\downarrow}l_{\bf p\downarrow})
+ S^+_{\bf n}r^\dagger_{\bf k\downarrow}l_{\bf p\uparrow}
+ S^-_{\bf n}r^\dagger_{{\bf k}\uparrow}l_{\bf p\downarrow}\right] + h.c.,
\nonumber\\
{\cal H}_T^{ep} &=&
\sum_{\alpha a \bf n,k,p} 
T^{ep,\alpha}_{\bf k,p}({\bf q})
r^\dagger_{{\bf k}a}l_{{\bf p}a}(b_{{\bf q} \alpha }
- b^\dagger_{{-\bf q}\alpha}) +h.c.
\label{eq:HT}
\end{eqnarray}
Here the surface $\Sigma$ lies somewhere in the barrier and separates
the electrodes,
we have subtracted an average spin $S^3_{\bf n} - 
\langle S^3_{\bf n}\rangle$
in each of electrodes as part of the exchange potential,
the exchange vertex is $T^J \sim J_{\bf n}\exp(-\kappa w)$,
and the phonon vertex is related to the deformation potential $D$ 
in the usual way [$ T^{ep}(q) \sim \imath 
Dq(\hbar/2M\omega_q)^{1/2}\exp(-\kappa w)$],
where $M$ is the atomic mass,
${\bf q}$ is the phonon momentum, ${\bf n}$ marks the lattice sites,
 and the vertices contain
the square root of the barrier transparency.\cite{mahan,amb_tunn2}
The operators $l_a$ and $r_a$ annihilate electrons with spin $a$ on 
the left and right electrodes, respectively.
Two more things to note: (i) the summations over ${\bf p}$ and ${\bf k}$
always include density of initial $g_{La}$ and final $g_{Rb}$ 
states, that makes
an exchange {\em and} phonon contribution spin-dependent,
(ii) when the magnetic moments on the electrodes are at a mutual angle $\theta$,
one has to express the operator $r$ w.r.t. the lab system and then use it
in  ${\cal H}_T$ ({\ref{eq:HT}).

The tunnel current will be calculated within the linear response 
formalism as
\cite{mahan}
\begin{equation}
I(V,t) = {\imath e\over \hbar}\int^t_{-\infty} dt' 
\langle\left[dN_L(t)/dt, {\cal H}_T(t')\right]\rangle_0,
\label{eq:lresponse}
\end{equation}
where $N_L(t)=\sum_{{\bf p}a}l^\dagger_{{\bf p}a}(t)l_{{\bf p}a}(t)$
is the operator of the number of electrons on the left terminal in
the interaction representation, $\langle~~\rangle_0$ stands for the
average over ${\cal H}_0$,
$$
{\cal H}_T(t) = \exp(-\imath e V t/\hbar)A(t)+h.c.,~~~~~
A(t) = \sum_{{\bf p}a,{\bf k}b} T_{{\bf p}a,{\bf k}b}(t)
r^\dagger_{{\bf k}b}(t)l_{{\bf p}a}(t),
$$
the tunnel vertex $T$ is derived for each term in (\ref{eq:HT}),
and $V$ is the bias.
We shall later consider impurity-assisted tunneling within the
same general approach.

\begin{flushleft}
\underline{Elastic tunneling}
\end{flushleft}
We are now in position to calculate all contributions to the tunneling
current, the simplest being direct elastic tunneling due to ${\cal H}^0_T$.
It is worth noting that it can also be calculated 
from the transmission probabilities of electrons with spin $a$, 
$T_a = \sum_{b} T_{ab}$, is the transmission
probability, which has a particularly simple form for a square barrier
and $collinear$ [parallel (P) or antiparallel (AP)] moments
on the electrodes\cite{amb_tunn1}
We obtain the  following expression for 
the direct tunneling conductance, assuming  $m_1 = m_3$ (below the effective
mass in the barrier will be measured in units of $m_1$):
\begin{eqnarray}
{G^0\over A} &=&{1\over A}\left({I\over V}\right)_{V\rightarrow 0} =
G^0_{\rm FBF}( 1 + P_{\rm FB}^2 \cos(\theta) ),
\label{eq:Gdir}\\
G^0_{\rm FBF} &=& {e^2\over \pi\hbar} {\kappa_0\over \pi w}
\left[ { {m_2 \kappa_0(k_\uparrow + k_\downarrow)(\kappa_0^2
+m_2^2k_\uparrow
k_\downarrow)}
\over { (\kappa_0^2+m_2^2k_\uparrow^2)(\kappa_0^2 +
m_2^2k_\downarrow^2)}}
\right]^2 
e^{-2\kappa_0w},\hspace{.2in}{\rm and}
\label{eq:Gcorr}\\
P_{\rm FB} &=& { {k_\uparrow-k_\downarrow}\over{k_\uparrow +k_\downarrow}}~
{{\kappa_0^2-m_2^2k_\uparrow k_\downarrow}\over
{\kappa_0^2 + m_2^2k_\uparrow k_\downarrow}},
\end{eqnarray}
where $P_{\rm FB}$ is the effective polarization of the 
ferromagnetic (F) electrode in the presence of the barrier (B),
$\kappa_0 = [2m_2(U_0-E)/\hbar^2]^{1/2}$, and $U_0$ is the top of the
barrier.  
Eq.~(\ref{eq:Gdir}) corrects an expression derived
earlier \cite{slon} for the effective mass of the carriers in the
barrier.
By taking a typical value of $G/A =$4-5 $\Omega^{-1} 
{\rm cm}^{-2}$ (Ref.~\cite{moodera})
$k_\uparrow=1.09 \AA^{-1}$, $k_\downarrow=0.42 \AA^{-1}$, 
$m_1\approx 1$ (for itinerant $d$ electrons in Fe)\cite{stearns}
and a typical barrier height for Al$_2$O$_3$ 
(measured from the Fermi level $\mu$)
$\phi=U_0-\mu=3 {\rm eV}$,  and the thickness $w \approx 20~\AA$, one arrives at 
the following estimate for the effective mass in
the barrier:  $m_2 \approx 0.4$.\cite{shu} 
These values give the renormalized polarization 
$P_{\rm FeB} = 0.28$, down from the bulk value for iron
$P_{\rm Fe}= 0.4$ (Ref.\cite{tedrow,moodera}) 
Note that neglect\cite{slon} of the mass correction would make 
$P_{\rm FeB}<0$, and this is not corroborated by experimental evidence.

In the standard  approximation of a rectangular shape the barrier
height is
$U_0=\frac{1}{2}(\phi_L + \phi_R - eV)$ and this leads to a quick rise
of the conductance with bias, $G^0(V)= G^0+const\cdot V^2$ at small $V$
($\phi_L$ and $\phi_R$ are the work functions of the electrodes).
In practice, the barrier parameters should be extracted from independent
experiments, such as internal photoemission, etc., but here we are
concerned with the generic behavior, where the present formalism is sufficient
for qualitative and even semi-quantitative analysis.
Since the barrier shape depends in a non-trivial manner on image forces,
the calculations have been performed numerically with actual barrier shape
at finite temperatures (Fig.~1).

We note that the (undesirable) downward renormalization of the
polarization rapidly goes with diminishing effective carrier mass
in the barrier. The renormalization is completely absent in 
half-metallic ferromagnets with Re$k_\downarrow=0$, as we shall
discuss below.

We define the magnetoresistance as the relative change in contact
conductance with respect to the change of mutual orientation of spins
from parallel ($G^{\rm P}$ for $\theta=0$)  to antiparallel ($G^{\rm AP}$ for
$\theta=180^\circ$)
as 
\begin{equation}
 MR = (G^{\rm P}-G^{\rm AP})/G^{\rm AP}=2P_{\rm FB}P'_{\rm
 FB}/(1-P_{\rm FB}P'_{\rm FB}).
\label{eq:MR}
\end{equation}
The most striking feature of Eqs.~(3),(4) is that the $MR$
tends to infinity for vanishing ${\rm Re}k_\downarrow$,
i.e. when both electrodes are made of a 100\%
spin-polarized material ($P=P'=1$), because of a gap in the density of
states (DOS) for minority
carriers up to their conduction band minimum $E_{CB\downarrow}$.
Then $G^{\rm AP}$ vanishes together with 
the tunnel probability, since there is a zero DOS
at $E=\mu$ for both spin directions.

Such half-metallic behavior is rare,
but some materials possess this amazing property, most interestingly the
oxides CrO$_2$ and Fe$_3$O$_4$.\cite{kats} These oxides have potential
for future applications in combination with lattice-matching materials, as we
shall illustrate below.

A more accurate analysis of the $I-V$ curve requires a numerical
evaluation of the tunnel current for arbitrary biases and 
image forces, and the results are shown in Fig.~1.
The top panel in Fig.~\ref{fig:fig1} 
shows $I-V$ curves for an iron-based F-B-F junction
with the above-mentioned parameters.
The value of TMR is about 30\% at low biases and steadily decreases with
increased bias. In a half-metallic case (Re$k_\downarrow=0$,
Fig.~\ref{fig:fig1}, middle panel, where a threshold
$eV_c=E_{CB\downarrow}-\mu = 0.3$~eV has been assumed),
we obtain {\em zero} conductance $G^{AP}$ in the AP 
configuration at biases lower than $V_c$.
It is
easy to see that above this threshold,
$G^{AP} \propto (V-V_c)^{5/2}$ at temperatures much smaller than 
$eV_c$.\cite{amb_tunn1}
 Thus, for $|V|<V_c$ in the AP geometry one has $MR=\infty$. 
In practice, there are several
effects that reduce this MR to some finite value, notably an
imperfect AP alignment of moments in the electrodes. 
However, from the middle and the bottom panels in Fig.~\ref{fig:fig1}
 we see that 
even at 20$^\circ$ deviation from the AP configuration, the value of MR 
 exceeds 3,000\% within the half-metallic gap $|V|<V_c$, 
and this is indeed a very large value.
%
\begin{figure}[t]
\epsfxsize=3.3in
\epsffile{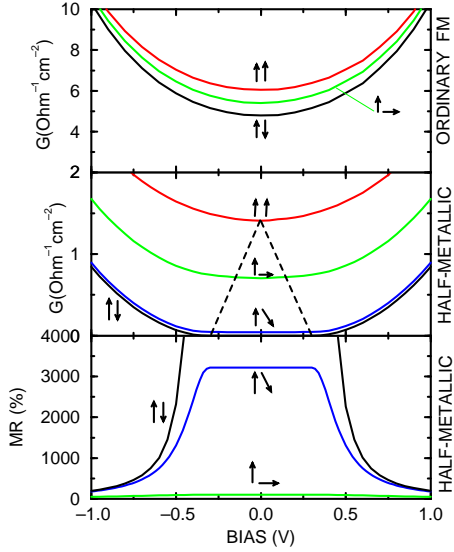}
\caption{\footnotesize
Conductance and magnetoresistance of tunnel junctions versus bias. 
Top panel: conventional
(Fe-based) tunnel junction (for parameters see text). 
Middle panel: half-metallic electrodes. Bottom panel:
magnetoresistance for
the half-metallic electrodes. Dashed line shows schematically a
region where a half-metallic gap in the minority spin states is controlling the transport.
Even for imperfect antiparallel alignment ($\theta=160^\circ$,
marked $\uparrow\searrow$), 
the magnetoresistance for half-metallics (bottom panel) 
exceeds 3000\% at biases below the threshold $V_c$.
All calculations have been performed at
 300K  with the inclusion of
multiple image potential and exact transmission coefficients.
Parameters are described in the text.  
\label{fig:fig1}
}
\end{figure}

\begin{flushleft}
\underline{Impurity-assisted tunneling}
\end{flushleft}
An important aspect of spin-tunneling is the effect of tunneling 
through the defect
states in the (amorphous) oxide barrier. Since the devices under consideration
are very thin, their $I-V$ curves and MR should be very sensitive to 
defect resonant states in the barrier with energies close to the chemical
potential, forming ``channels'' with the nearly periodic positions of
impurities (Fig.~\ref{fig:impur}).\cite{larkin}
Generally, channels with one impurity (most likely to dominate in 
thin barriers) would result in a monotonous behavior of the $I-V$ curve,
whereas channels with {\em two or more} impurities would produce
intervals with negative differential conductance.\cite{larkin}

Impurity-assisted spin tunneling  at zero temperature
(at non-zero T one should include an integration with the Fermi 
functions) has a resonant form\cite{larkin,amb_tunn1}
\begin{equation}
G_a  = {2e^2 \over \pi\hbar}
\sum_i {\Gamma_{La}\Gamma_{Ra} \over{(E_i-\mu)^2+\Gamma^2}},
\label{eq:breit}
\end{equation}
where $\Gamma_a = \Gamma_{La} + \Gamma_{Ra}$ 
is the total width of the resonance
given by the sum of the partial widths $\Gamma_L$ ($\Gamma_R$) corresponding
to electron tunneling from the impurity state at the energy $E_i$ to 
the left (right) terminal. 
For the tunnel width we have 
\begin{equation}
\Gamma_{(L,R)a}~=~2\pi^2 \kappa_0 (\hbar^2/
m_2 )^2 \sum_{ {\bf k}_{(L,R)a} } |\psi_{{\bf k}_{(L,R)a}
}({\bf n}_i)|^2 \delta(E_{\bf k} - E_i),
\label{eq:Gamma_i}
\end{equation}
where $\psi_{ {\bf k}_{(L,R)a}}({\bf n}_i)$ is the value of the electrode
wave function, exponentially decaying into the barrier, at an impurity site
$ {\bf n}_i$. For a rectangular barrier 
we have \cite{amb_tunn1}
\begin{equation}
\Gamma_{La} = \epsilon_i~ {2m_2 k_a \over
{ \kappa_0^2 + m_2^2k_a^2 }}~
{e^{-\kappa_0(w+2z_i)}\over {\kappa_0 (\frac{1}{2}w + z_i)}},
\label{eq:Gsb}
\end{equation}
where $z_i$ is the coordinate of the impurity with respect to the center of the
barrier,
$\epsilon_i= \hbar^2 \kappa_0^2/(2m_2)$.
For e.g. P configuration and electrodes of the same material,
the conductance would then be proportional to
$\left[(E_i-\mu)^2+4\Gamma^2_{0a}\cosh^2(2\kappa_0 z_i)\right]^{-1}$,
where $\Gamma_{0a}$ equals (\ref{eq:Gsb}) without the factor
$\exp(-2\kappa_0 z_i)$ [c.f. Eq.~(\ref{eq:N1})].
The conductance has a sharp maximum ($=e^2/(2\pi \hbar)$) when $\mu = E_i$
and $\Gamma_L=\Gamma_R$, i.e. for the symmetric position of the impurity in
the barrier $|z_i|<1/\kappa_0$ in a narrow interval of energies $|\mu-E_i|<\Gamma$.
Averaging over energies and positions of impurities in Eq.~(\ref{eq:breit}),
and considering a general configuration of the magnetic moments on the
terminals, we get the following formula for impurity-assisted conductance
in the leading order in $\exp(-\kappa w)$:
\begin{equation}
{G^1\over A} = G^1_{\rm imp} (1+\Pi_{\rm FB}^2 \cos(\theta)),
\label{eq:Gimp}
\end{equation}
where we have introduced the quantities
\begin{eqnarray}
G^1_{\rm imp} &=& {e^2\over \pi\hbar}N_1, \hspace{.2in} N_1 = \pi^2 \nu
\Gamma_1/\kappa_0,\nonumber\\
\Gamma_1 &=& \epsilon_i{e^{-\kappa_0 w}\over\kappa_0w}
\left(r_\uparrow + r_\downarrow \right)^2, \hspace{.2in}
\Pi_{\rm FB} = (r_\uparrow - r_\downarrow)/(r_\uparrow +
r_\downarrow),\hspace{.2in}{\rm and}\nonumber\\
r_a &=& [m_2\kappa_0 k_a/(\kappa_0^2 +  m_2^2 k_a^2 )]^{1/2},
\label{eq:N1}
\end{eqnarray}
with $N_1$ being the effective number of one-impurity channels 
per unit area,
and $\Pi_{\rm FB}$ is the `polarization' of the impurity 
channels. When the total number of one-impurity channels ${\cal
N}_1=N_1A\gg 1$, then we will have a self-averaged conductance,
otherwise the conductance will depend on a specific arrangement
of impurities (regime of mesoscopic fluctuations).

Comparing the direct and the impurity-assisted contributions to
conductance, we 
see that the latter dominates when the impurity density of states
$\nu 
\geq (\kappa_0/\pi)^3
\epsilon_i^{-1}\exp(-\kappa_0w)$,
and in our example a  crossover takes place at $\nu 
\geq 10^{17}$cm$^{-3}$eV$^{-1}$. When the resonant transmission 
dominates, the magnetoresistance is given by
\begin{equation}
MR_1=2\Pi~\Pi'/(1-\Pi~\Pi'),
\label{eq:mr1}
\end{equation}
which is just 4\% in the case of Fe. Thus, we have a drastic reduction
of the TMR due to non-magnetic impurities in the tunnel barrier,
and in the case of magnetic impurities the TMR will be even smaller.

With standard ferromagnetic 
electrodes, 
the conductance is exponentially enhanced [$G^1 \propto \exp(-\kappa_0w)$,
whereas $G^0   \propto \exp(-2\kappa_0w)$] but the magnetoresistance is reduced
in comparison with the `clean' case of a low concentration of defect levels.
These predictions\cite{amb_tunn1} have been confirmed by recent
experiments.\cite{janice,mood_def}

With further increase of the defect density and/or the barrier width,
the channels with two- and more impurities will become more effective
than one-impurity channels described above, as has been known for 
quite a while.\cite{pollak,larkin}   
The contribution of
the many-impurity channels, generally, will result in
the appearance of irregular intervals with negative differential
conductance on the $I-V$ curve.\cite{larkin}
Thus, the two-impurity channels define random fluctuations of
current with bias.
This is due to the fact that the energy of defect states depends on
bias as $\epsilon_i = \epsilon_i^0 + eVz/w$. With increasing 
bias (i) the total number of two-impurity channels increases
but (ii) some of these channel go off resonance and reduce their
conductance. Accidentally, the
number of two-impurity channels going off resonance may become larger
than a number of new channels, leading to suppressed overall
conductance.
If we denote by $\Gamma_2$ the width of the two-impurity channels,
then the fluctuations would obviously  occur on a scale 
$\Delta V < \Gamma_2/e$. Then, according to standard arguments,
the change in current will be
\begin{equation}
{\Delta I \over I} = {\Delta V\over V} \pm \left({e\Delta V\over 
\Gamma_2}\right){\cal N}^{-1/2},
\label{eq:dI}
\end{equation}
where ${\cal N}=eV{\cal N}_2/\Gamma_2$ is the number of the two-impurity
channels contributing at the bias $V>\Gamma_2/e$, 
${\cal N}_2$ is the total number of the two-impurity channels,
${\cal N}_2=A\pi^3w^3\nu^2\Gamma_2^2\kappa_0^{-1}$, and
$\Gamma_2=(4\epsilon_i\Gamma_1/
(\kappa_0 w))^{1/2}$.\cite{larkin} When $eV/\Gamma_2 > 
{\cal N}_2(\kappa_0 w)^2$,
then the second (random) term in (\ref{eq:dI}) exceeds the first term,
and this leads to random intervals with negative differential
conductance.
Obviously, with increasing temperature or/and bias in thick enough
barriers, longer and longer impurity channels will be `turned on'.
A corresponding  microscopic model should include impurity states
coupled to a phonon bath, and such a model has been solved in 
Ref.~\cite{glazman}.
The authors found an average conductance due to an $n$-impurity chain
in the limit $eV \ll T$, which gives
for $n=2$~~  $G_2(T) \propto T^{4/3}$.
In the opposite limiting case of $eV\gg T$, 
the result is:\cite{glazman} $G_2(V)
\propto  V^{4/3}$, and this crossover behavior is indeed in very good
agreement with experiments on a-Si barriers.\cite{beasley}

One may try to fabricate a resonant tunnel diode (RTD) structure to sharply
increase the conductance of a system. 
We can imagine an RTD structure with an extra thin non-magnetic layer
placed between two oxide barrier layers producing a resonant level at
some energy $E_r$. The only difference from the previous discussion is 
the effectively $1D$ character of the transport in the RTD in comparison with $3D$ 
impurity-assisted transport. 
However, the transmittance will have the same 
resonant form as in (\ref{eq:breit}) and the widths (\ref{eq:Gsb}).
The estimated magnetoresistance in
the RTD geometry  is, with the use of (\ref{eq:breit}),
\begin{equation}
MR_{\rm RTD} = \left[ (r^2_\uparrow  -  r^2_\downarrow)/
( 2 r_\uparrow r_\downarrow)\right]^2,
\end{equation}
which is about 8\% for Fe electrodes. We see that the presence of random
 impurity levels
 or a single resonant level reduces the value of the magnetoresistance as
compared with direct tunneling.\\

\noindent
\underline{Roughness}\\
As we have seen, the conductance is dominated by the
exponentially small barrier transparency,
$\propto \exp(-2w(\kappa_0^2+{\bf k_\parallel}^2)^{1/2})$, 
so that the 
contribution comes mainly from electrons tunneling perpendicular to
the barrier, i.e. with small parallel momenta
$|{\bf k_\parallel}|<(\kappa_0/w)^{1/2}$. For barriers with
a rough interface $w=\overline{w}+h$, where $h$ is the height of
asperities  and $\overline{w}$ is the average barrier thickness.
Each asperity will contribute a factor of
$\exp(2\kappa_0 h)$ to the conductance,
which we have to average.
We assume a normal distribution for roughness, $P(h)=(2\pi h_0^2)^{-1/2}
\exp\left(-h^2/(2h_0^2)\right)$. Then, the average conductance $\overline{G}$
becomes
\begin{equation}
G = \overline{G}\int^\infty_{-\infty} dh\exp(2\kappa_0 h)P(h) =
\overline{G}\exp(\kappa_0^2 h_0^2)\propto
\exp[-2\kappa_0(\overline{w}-\kappa h_0^2)].
\label{eq:rough}
\end{equation}
This result means that the effective thickness of the barrier is
reduced by $\kappa h_0^2$ in comparison with the observed average 
thickness $\overline{w}$. 
The generalization for the case of correlated roughness is straightforward
and does not change this result.\\
\begin{figure}[t]
\epsfxsize=2.4in
\epsffile{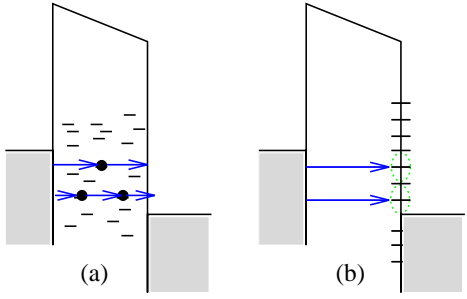}
\caption{\footnotesize
Schematic of tunneling via chains of the localized states in the barrier (a)
and into the localized surface states (b).
\label{fig:impur}
}
\end{figure}

\noindent
\begin{flushleft}
\underline{ Tunneling via Surface States}
\end{flushleft}
Direct tunneling, as we have seen, gives TMR of about 30\%,
whereas the recent experimental 
results are almost ten percent higher.\cite{janice,mood_zb}
As we shall see shortly, this moderate difference is unlikely to come
from the inelastic processes. Up to now we have disregarded
the possibility of localized states at metal-oxide interfaces.
Keeping in mind that the usual barrier AlO$_x$ is amorphous,
the density of such states may well exceed that at typical 
semiconductor-oxide surfaces. If this is true, then
we have to take into account tunneling into/from those states.
 If we assume that electrons at the surface are confined by a
short-range potential then we can estimate the tunneling matrix
elements as described above.
  The corresponding tunneling MR is given by 
\begin{eqnarray}
{G_{\rm bs}(\theta)\over A} &=& {e^2\over \pi\hbar} B\overline{D}_s(1+P_{\rm
FB}P_s\cos(\theta)),\nonumber\\
P_s &=& {D_{s\uparrow} - D_{s\downarrow}\over{D_{s\uparrow} 
+ D_{s\downarrow}}},\hspace{.2in}\nonumber\\
\overline{D}_s&=&\frac{1}{2}
(D_{s\uparrow}+D_{s\downarrow}),\nonumber\\
B&=&8\pi^2{\epsilon_s\over \kappa_0 w}
{ {m_2\kappa_0(k_\uparrow + k_\downarrow)(\kappa_0^2
+m_2^2k_\uparrow
k_\downarrow)}
\over { (\kappa_0^2+m_2^2k_\uparrow^2)(\kappa_0^2 +
m_2^2k_\downarrow^2)}}
\exp(-2\kappa_0 w),
\label{eq:Gsurface}
\end{eqnarray}
where $P_s$ is the polarization and $\overline{D}_s$
is the average density of surface states, 
$\epsilon_s = \hbar^2\kappa_0^2/(2m_2)$.
The corresponding magnetoresistance would be 
$MR_{\rm bs} = 2P_{\rm FB}P_s/(1-P_{\rm FB}P_s)$.

Comparing (\ref{eq:Gsurface})
with (\ref{eq:Gdir}), we see that the bulk-to-surface
conductance exceeds bulk-to-bulk tunneling at moderate
densities of surface states $D_s>D_{sc}\sim 10^{13}$cm$^{-2}$eV$^{-1}$
per spin.

If on both sides the density of surface states is above critical value
$D_{sc}$,
the magnetoresistance will be due to surface-to-surface  tunneling
with a value given by
$$
MR_{\rm ss} = 2P_{s1}P_{s2}/(1-P_{s1}P_{s2}),
$$
and if the polarization of surface states is larger than in the bulk,
then it would result in enhanced TMR.
This mechanism may be even more relevant for Fe/Si and other
ferromagnet-semiconductor structures.\cite{alison}
\\

\noindent 
INELASTIC TUNNELING, `ZERO-BIAS' ANOMALY\\

So far we have disregarded all inelastic processes, such as phonon
emission by the tunneling electrons. These processes were long thought to be
responsible for a so-called `zero-bias' anomaly observed in a variety of
non-magnetic\cite{duke} and magnetic junctions.\cite{janice,mood_zb}
Magnetism in electrodes introduces
new peculiarities into the problem, which we will now discuss.
The obvious one is related to emission of magnons.
At temperatures well below the Curie temperature and not very large biases,
one can describe spin excitations
by introducing magnons. Then the calculations of exchange- and 
phonon-assisted currents become very similar. Thus, we obtain from
(\ref{eq:lresponse}) and (\ref{eq:HT}) the following expression for
magnon-assisted current in e.g. parallel configuration (corresponding
expressions can be easily found for other configurations as  well):
\begin{eqnarray}
I^x_{\rm P}(V,T) &=& 
{2\pi e\over \hbar}\sum_{{\bf q}\alpha}X^\alpha
\biggl(
g^L_\uparrow g^R_\downarrow ~(eV+\omega)
\left[
{N_\omega\over {1-\exp(-\beta(eV+\omega))}}
+{{N_\omega+1}\over{ 1-\exp(\beta(eV+\omega))}}
\right]\nonumber\\
&+&g^L_\downarrow g^R_\uparrow ~(eV-\omega)
\left[
{{N_\omega+1}\over{1-\exp(-\beta(eV-\omega))}}
+{N_\omega\over{1-\exp(\beta(eV-\omega))}}\right]
\biggr),
\label{eq:IxP}
\end{eqnarray}
where $N_\omega = \left[\exp(\beta\omega)-1\right]^{-1}$,
$\beta=1/T$ is the inverse temperature, 
$\omega = \omega^\alpha_{\bf q}$ and 
$X^\alpha$ is the magnon incoherent vertex
related to the $|T^{x,\alpha}_{\bf p,k}(2S_{\bf n}/N)^{1/2}|^2$ 
(\ref{eq:HT})
with all momenta parallel to the barrier integrated out.\cite{amb_tunn2}
To get this expression, we have also assumed that the electron
densities of states  $g$ in (\ref{eq:IxP}) 
vary on a larger scale than the bosonic contributions do,
and, therefore, substituted them by representative values at the nominal
Fermi levels. If there are some fine features in the electron DOS, then
the integral over electron energies should remain, thus
necessarily smoothing out any such fine features in the electron DOS.
\begin{figure}[t]
\epsfxsize=3.in
\epsffile{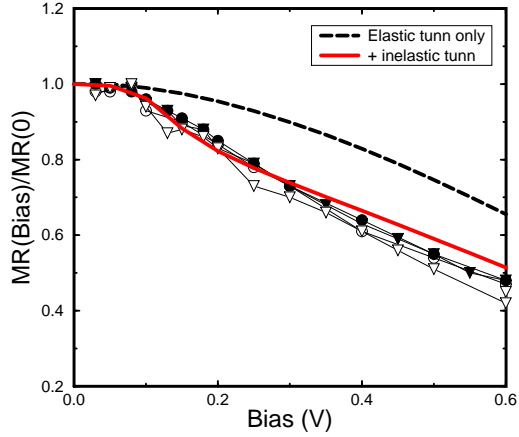}
\caption{\footnotesize
Fit to experimental data for the magnetoresistance
of Co/Al$_2$O$_3$/NiFe tunnel junctions [12] 
with inclusion of elastic and inelastic (magnons and phonons)
tunneling. 
The fit gives for magnon DOS $\propto \omega^{0.65}$ which is 
close to the standard spectrum  $\propto \omega^{1/2}$.
\label{fig:fit}
}
\end{figure}

For the limiting case of $T=0$, we obtain for inelastic tunneling current:
\begin{eqnarray}
I^x_{\rm P} &=& {2\pi e\over \hbar}\sum_\alpha X^\alpha g^L_\downarrow g^R_\uparrow
\int d\omega 
\rho^{mag}_{\alpha}(\omega) (eV-\omega)\theta(eV-\omega),\nonumber\\
I^x_{\rm AP} &=& {2\pi e\over \hbar}\biggl[
X^R g^L_\uparrow g^R_\uparrow
\int d\omega \rho^{mag}_{R}(\omega) (eV-\omega)\theta(eV-\omega)\nonumber\\
&+&
X^L g^L_\downarrow g^R_\downarrow
\int d\omega \rho^{mag}_{L}(\omega) (eV-\omega)\theta(eV-\omega)\biggr].
\label{eq:Ix0}
\end{eqnarray}
where $\theta(x)$ is the step function,
$\rho_\alpha^{mag}(\omega)$ is the magnon density of states
that has a
general form $\rho_\alpha^{mag}(\omega)
=(\nu+1)\omega^{\nu}/\omega_0^{\nu+1}$, $\nu$ can be used as a fitting
parameter to define a dispersion of the relevant magnons, and
$\omega_0$ is the maximum magnon frequency. For phonon-assisted
current at $T=0$ we have 
\begin{eqnarray}
I^{ph}_{\rm P} &=& {2\pi e\over \hbar}\sum_{a\alpha} g^L_a g^R_a
\int d\omega  \rho^{ph}_\alpha(\omega) P^\alpha(\omega)(eV-\omega)\theta(eV-\omega),\\
I^{ph}_{\rm AP} &=& {2\pi e\over \hbar}\sum_{a\alpha} g^L_a g^R_{-a}
\int d\omega  \rho^{ph}_\alpha(\omega) P^\alpha(\omega)(eV-\omega)\theta(eV-\omega).
\label{eq:Ip0}
\end{eqnarray}
One can show that the ratio of phonon to exchange vertex is
$P(\omega)/X = \gamma \omega/\omega_D$, where $\gamma$ is a constant
depending on the ratio between deformation potential and 
exchange constants,\cite{amb_tunn2} and $\omega_D$ is the Debye frequency.

The elastic and inelastic contributions together will define the total
junction conductance $G=G(V,T)$ as a function of the bias $V$ and
temperature $T$.
We find that the inelastic contributions from magnons and phonons
(\ref{eq:Ix0})-(\ref{eq:Ip0}) grow as 
$G^x(V,0) \propto (|eV|/\omega_0)^{\nu+1}$ and 
$G^{ph}(V,0) \propto (|eV|/\omega_D)^4$
at low biases. 
These contributions saturate at higher biases: $G^x(V,0) \propto
1-\frac{\nu+1}{\nu+2}\frac{\omega_0}{|eV|}$ at $|eV|>\omega_0$;
$G^{ph}(V,0) \propto 
1-\frac{4}{5}\frac{\omega_D}{|eV|}$ at $|eV|>\omega_D$.
This behavior would lead to sharp features in the
$I-V$ curves on a scale of 30-100 mV (Fig.~\ref{fig:fit}).

It is important to highlight the opposite effects of phonons and magnons
on the TMR. If we take the case of the same electrode materials and
denote $D=g_\uparrow$ and $d=g_\downarrow$ then we see that
$G^x_{\rm P}(V,0) - G^x_{\rm AP}(V,0) 
\propto - (D-d)^2(|eV|/\omega_0)^{\nu+1}<0$, whereas
$G^{ph}_{\rm P}(V,0) - G^{ph}_{\rm AP}(V,0) 
\propto +(D-d)^2(|eV|/\omega_D)^4>0 $, i.e.
spin-mixing due to magnons {\em kills}, whereas the phonons tend to
{\em enhance} the TMR.\cite{zhang}

Finite temperature gives contributions
of the same respective sign as written above.
For magnons: 
$G^x_{\rm P}(0,T) - G^x_{\rm AP}(0,T) 
\propto - (D-d)^2 (-TdM/dT)<0$,
where $M=M(T)$ is the magnetic moment of electrode at given 
temperature $T$. Phonon contribution is given by standard Debye integral
with the following results:
$G^{ph}_{\rm P}(0,T) - G^{ph}_{\rm AP}(0,T) 
\propto +(D-d)^2(T/\omega_D)^4>0 $ at $T\ll\omega_D$, and 
linear temperature dependence at high temperatures
$G^{ph}_{\rm P}(0,T) - G^{ph}_{\rm AP}(0,T) \propto
+(D-d)^2(T/\omega_D)$
 at $T\gg \omega_D$.\cite{amb_tunn2} 
We note again an opposite effect of magnons
and phonons on the tunneling magnetoresistance.

We have not included Kondo\cite{appel} and other correlation 
effects that might contribute at very low biases, since they usually 
do not help to quantitatively fit the data.\cite{beasley}

The role of phonons is illustrated by my fit to recent experiments
carried out at HPL:\cite{janice} it appears that only after including
phonons is it possible to get a sensible fit to the magnon DOS 
with $\nu=0.65$, which is close to the bulk value $\frac{1}{2}$ 
and $\gamma \approx 0.1$ (Fig.~\ref{fig:fit}).

\noindent
\begin{flushleft}
{100\% POLARIZATION}
\end{flushleft}
It is very important that {\em in the case of half-metallics}
$r_\downarrow=0$, $\Pi_{\rm FB}=1$,
and even with an imperfect barrier magnetoresistance can, 
at least in principle, reach any value
limited by only spin-flip processes in the barrier/interface and/or 
misalignment of moments in the half-metallic ferromagnetic 
electrodes.\cite{amb_tunn1}
We should note that the {\em one-magnon} excitations in
half-metallics are suppressed by the half-metallic gap, as
immediately follows from our discussion in the previous section.
Spin-mixing can only occur on magnetic impurities
in the barrier or interface, because the allowed {\em two}-magnon
excitations in the electrodes do not result in spin-mixing.

Therefore, these materials should combine the best of both worlds: 
very large magnetoresistance with enhanced
conductance in tunnel MR junctions.
One should be aware, however, that defect states (like unpaired
electrons) will increase the spin-flip rate, so the magnetoresistance 
could vanish with an increasing concentration of defects.
In the case of conventional systems (e.g. FeNi electrodes) we have seen,
however, that resonant tunneling significantly reduces the tunnel MR
by itself, so the possibility of improving the conductance and still having
a very large magnetoresistance resides primarily with half-metallics.
\begin{figure}[t]
\epsfxsize=3.4in
\epsffile{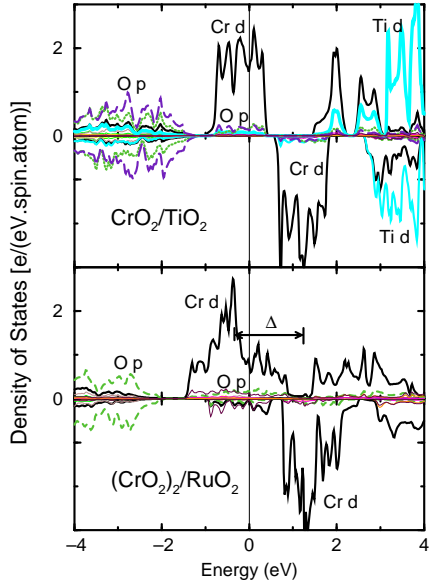}
\caption{\footnotesize
Density of states of
CrO$_2$/TiO$_2$ (top panel)
and (CrO$_2$)$_2$/RuO$_2$ (bottom panel) half-metallic 
layered structures calculated with the
use of the LMTO method.
\label{fig:cro2}
}
\end{figure}
I shall finish with a couple of examples of systems 
with half-metallic behavior, CrO$_2$/TiO$_2$ and 
CrO$_2$/RuO$_2$\cite{amb_tunn1} (Fig.~\ref{fig:cro2}).
They are based on half-metallic CrO$_2$, and all materials 
have the rutile structure with almost perfect lattice 
matching, which should yield a good interface
and should help in keeping the system at the desired stoichiometry.
TiO$_2$ and RuO$_2$ are used as the barrier/spacer oxides.
The half-metallic behavior of the corresponding 
multilayered systems is  demonstrated by the
band structures calculated within the linear muffin-tin orbitals method (LMTO)
in a supercell geometry with [001] growth
direction and periodic boundary conditions.
The calculations show that CrO$_2$/TiO$_2$ is a perfect half-metallic, whereas
(CrO$_2$)$_2$/RuO$_2$ is a weak half-metallic, since 
there is some small DOS around $E_F$, and an exact gap opens up
at about 0.58 eV above the Fermi level (Fig.~\ref{fig:cro2}).
In comparison, 
there are only states in the majority spin band at the Fermi level
in CrO$_2$/TiO$_2$. An immediate consequence of the fact that minority
spin bands are fully occupied is an exact {\em integer} 
value of the magnetic moment in the unit cell (=2$\mu_B$/Cr in
CrO$_2$/TiO$_2$), and  
this property is a simple check for possible {\em new} half-metallics.

The electronic structure of CrO$_2$/TiO$_2$ is very
interesting in that it has a half-metallic gap which is 2.6 eV wide
and extends on both sides of the Fermi level, where there is a gap
either in the minority {\em or} majority spin band. Thus, 
an huge magnetoresistance should in principle be seen not only for
electrons at the Fermi level biased up to 0.5 eV, but also for {\em hot}
electrons starting at about 0.5 eV above the Fermi level.
We note
that states at the Fermi level are a mixture of Cr($d$) and O($2p$) states,
so that $p-d$ interaction within the first coordination shell produces a
strong hybridization gap,  and  the  Stoner spin-splitting
moves the Fermi level right into the gap for minority carriers
(Fig.~\ref{fig:cro2}).

An important difference between the two spacer oxides 
is that TiO$_2$ is an insulator whereas RuO$_2$ is a good
metallic conductor. Thus, the former system can be used in a tunnel junction,
whereas the latter will form a metallic multilayer. In the latter case
the physics of conduction is different from tunneling but the effect
of vanishing phase volume for transmitted states still works when
current is passed through such a system {\em
perpendicular to planes}. For the P orientation 
of moments on the electrodes, CrO$_2$/RuO$_2$
would have a normal metallic conduction, whereas in the AP one
we expect it to have a semiconducting type of transport, with a
crossover between the two regimes.
One interesting possibility is to form  three-terminal devices with 
these systems, like a spin-valve transistor,\cite{monsma}
and check the effect in the hot-electron region.
CrO$_2$/TiO$_2$ seems to a be a natural candidate to check the present
predictions about half-metallic behavior and for a possible large
tunnel magnetoresistance. An important advantage of these systems is
almost perfect lattice matching at the oxide interfaces. 
The absence of such a match of the conventional Al$_2$O$_3$ barrier with
Heusler half-metallics (NiMnSb and PtMnSb) may have been among other
reasons for their moderate performance.\cite{hmfexp}

By using all-oxide half-metallic systems, 
as described herein, one may  bypass many  materials issues.
Then,  the main concerns for achieving a very
large value of magnetoresistance will be spin-flip centers
 and imperfect alignment of moments.
As for conventional tunnel junctions, the
present results show that the presence of defect states in the barrier, 
or a resonant state like in a resonant tunnel diode type of
structure, reduces their magnetoresistance by several times but may
dramatically increase the current through the structure.

Finally, we can mention the CMR materials.
Experiment\cite{cmr_opt} and LDA  calculations\cite{singh} 
indicate that manganites are close to half-metallic behavior
as a result of a significant spin-splitting presumably due to strong Hund's rule
coupling on Mn.
Manganites are strongly correlated materials, likely with electronic
phase separation,\cite{nagaev} which makes their study a real challenge.
There are a number of studies of systems, where transport
is going across grain boundaries or between MnO$_2$ layers in tailored
derivatives of the perovskite phase.\cite{cmr_tunn}
A hope is that some of these structures with manganites might
operate at low fields and reasonably high temperatures.\cite{hwang} 
The low field (below 1000 Oe) TMR
in polycrystalline La$_{2/3}$Sr$_{1/3}$MnO$_3$ perovskite and 
Tl$_2$Mn$_2$O$_7$ pyrochlore is about 30\% and is likely due to intergrain 
carrier transport.
It would be interesting to apply the results of
the present work to tunneling phenomena in the CMR-based 
layered/inhomogeneous structures.
For instance, CrO$_2$ junctions would help to check on
the relevance of the half-metallic behavior to conduction in the CMR
materials.
In particular, it should be signalled by
a plateau in the tunneling magnetoresistance as a function of bias
within the half-metallic band gap (Fig.~\ref{fig:fig1}).

\normalsize

%
I am grateful to J.~Nickel, T.~Anthony, J. Brug, and J. Moodera
for sharing their data, and
to G.A.D.~Briggs, N. Moll, and R.S.~Williams for 
useful discussions.

\begin{flushleft}
REFERENCES
\end{flushleft}

\end{document}